\documentclass[conference]{IEEEtran}
\IEEEoverridecommandlockouts
\usepackage[utf8]{inputenc}
\usepackage{cite}
\usepackage{amsmath,amssymb,amsfonts}
\usepackage{algorithmic}
\usepackage{graphicx}
\usepackage{textcomp}
\usepackage{xcolor}

\usepackage{latexsym}
\usepackage{mathptmx}
\usepackage{amsbsy}
\usepackage{amsthm}
\usepackage{array}
\usepackage{multirow}
\usepackage{hhline}		
\usepackage{subfigure}
\usepackage{mathptmx}
\usepackage{mathtools}
\usepackage{todonotes}

\usepackage{enumitem} 
\usepackage[square,sort&compress,comma,numbers]{natbib}

\usepackage{booktabs} % For formal tables
\usepackage{float}
\usepackage{scrhack}
\usepackage{listings}
\usepackage{color}
\usepackage{pgfplots}
\usepackage{tikz}
\usepackage{cleveref}
\usepackage{here}
\usepackage{url}
\usepackage{enumitem}
\usepackage{lipsum}
\usepackage{eqnarray}
\usepackage{breakurl}

\def\BibTeX{{\rm B\kern-.05em{\sc i\kern-.025em b}\kern-.08em
    T\kern-.1667em\lower.7ex\hbox{E}\kern-.125emX}}
\begin{document}

\title{Selective Deletion in a Blockchain}
%\title{Blockchain Technology with the Functionality of Oblivion}
%\title{Blockchain Deletion: a second life}
%\title{Enhancement of the blockchain with the forgetting functionality}

\author{\IEEEauthorblockN{Peter Hillmann, Marcus Kn\"upfer, Erik Heiland, and Andreas Karcher}
\textit{Bundeswehr University Munich}\\
Werner-Heisenberg-Weg 39, 85579 Neubiberg, Germany \\
Email: \{peter.hillmann, marcus.knuepfer, erik.heiland, andreas.karcher\}@unibw.de}
%\and
%\and
%\and
%\and
%\and

\maketitle

\begin{abstract}
The constantly growing size of blockchains becomes a challenge with the increasing usage.
Especially the storage of unwanted data in a blockchain is an issue, because it cannot be removed naturally.
In order to counteract this problem, we present the first concept for the selective deletion of single entries in a blockchain.
For this purpose, the general consensus algorithm is extended by the functionality of regularly creating summary blocks.
Previous data of the chain are summarized and stored again in a new block, leaving out unwanted information.
With a shifting marker of the \textit{Genesis Block}, data can be deleted from the beginning of a blockchain.
In this way, the technology of the blockchain becomes fully transactional.
The concept is independent of a specific block structure, network structure, or consensus algorithm.
Moreover, this functionality can be adapted to current blockchains to solve multiple problems related to scalability.
This approach enables the transfer of blockchain technology to further fields of application, among others in the area of Industry 4.0 and Product Life-cycle Management.
\end{abstract}


\section{Introduction}
The distributed ledger technology also known as the blockchain technology exists since 2008~\cite{Nakamoto2008}. 
It has become well-known through the crypto currencies Bitcoin and Ethereum.
Satoshi Nakamoto solved the Byzantine Generals Problem alongside the invention of Bitcoin, which is a common problem in distributed systems~\cite{Lamport1982}.
The Ethereum network has extended the blockchain technology by the Turing completeness in 2015.
It allows the execution of programs decentralized and to store so-called smart contracts in the blockchain.
In this way, it is possible to bind crypto tokens to material goods instead of trading virtual goods.
The increased public interest of the last years is still unbroken due to the high exchange rate of the stocks~\cite{buybitcoinworldwide2020}.
As a result more and more crypto currencies appeared.
Nevertheless, the technology of blockchain is also of interest in other areas like Industry 4.0 \cite{Laabs2018} and Supply Chain~\cite{Korpela2017}. %Authentification[6], logging, automotive, election
In Estonia, the blockchain is used at national level to administer health data~\cite{Einaste2018}.
The reason for this are the unique properties, whose transactions is desirable in many areas.
%The advantages of security, traceability and no need for third parties are decisive.
The advantages are decisive in terms of security, traceability and not requiring a third party.
The underlying technology of the blockchain provides a method for untrusted participants to reach a consensus over a public untrusted network.

The most important properties of a blockchain are the trustworthiness and reliability based on the immutability of the data.
By concatenating data sets in the form of blocks using the hash value, it is not possible to manipulate a block within the blockchain.
Thus the order as well as the content of the blocks are fixed.
To ensure the absolute correctness of a blockchain, it has to be completely available, starting with the \textit{Genesis Block}.
Therefore the blockchain is often called unhackable.
However, a change of the blockchain is possible, but becomes less likely with the increasing number of confirmation blocks.

Since the technology is still in the nascent stage, there are still problems with it:
\begin{itemize}
\item Blockchain attribute immutability in combination with illegal and unwanted content in a public blockchain
\item Growth of the blockchain
\item Laws with the right of erasure
\end{itemize}

The blockchain attribute immutability represents in these cases rather a problem than an advantage.
Researchers discovered that the Bitcoin blockchain contains pornographic content~\cite{Holland2018}.
This is possible because Bitcoin transactions allow a variety of non-Bitcoin related data to be stored alongside~\cite{Matzutt2018}.
%Since a full node has to download the complete blockchain to work in the network.
Since a full node must download the entire blockchain in order to work with the network, the illegal content is also downloaded and made available to others.
%So, the information about the illegal content is also downloaded and made available to others.
%In consequence of this is that some countries prohibit operating a full node or you are liable to prosecution.
As a result, some countries prohibit the operation of full nodes and make it liable to prosecution.

%This kind of problems led to a moral issue wheather to running a full node can be illegal.

Further, the growing size of a blockchain makes the operation very difficult in long term.
This problem becomes more intense as more people create transactions in a specific blockchain.
Bitcoin, released in 2009, has almost reached a blockchain size of 300 GB~\cite{Blockchair2020}.
Dependent on storage space and network bandwidth, the costs of operation increase significantly.
%The blockchain is a continuous sequence of data shaped into blocks that are cryptographically connected to their successor block. This ensures that entries cannot be manipulated afterwards. 

Last but not least, the right-to-erasure has to be respected, especially in case of personal data.
Currently, there is no clear answer on how to deal with such problems.

% Was wird in diesem Paper dargestellt? Was ist das Ziel? Was wird das Ergebnis sein?
With this paper, we present the first concept to delete unwanted data in a running blockchain.
The blockchain will forget this information instead of a direct deletion to keep a trusted chain.
This approach works independently of the network infrastructure and the consensus algorithm of a blockchain.
%Therefore, this topic is out of scope of this work.

%In this concept, we present a solution to delete stored data in a blockchain with elapsed time.
%So the intended action of deletion will take some steps to further guarantee the properties of a trusted blockchain.

The remainder of this paper is structured as follows.
In Section~II, we describe a simplified scenario to identify the requirements of the problem.
Section~III discusses related work in relation to the requirements.
The main part in Section~IV presents the concept of extending the functionality of a blockchain to be fully transactional.
Thereafter, Section~V shows some experiments and practical usage, before Section~VI concludes this paper.

\section{Scenario and requirements}\label{sec:scenario}
% The described use case gives rise to different requirements, which are explained in this section. The listed properties are used in the evaluation Chapter 7 to analyses the presented concept and to compare it with the methods presented in related work (see Chapter 4)
The General Data Protection Regulation (GDPR) was adopted in May 2016.
The objective of this directive is to standardize the handling of personal data of European citizens both within and outside of the European Union.
%and to make it more secure and transparent for data subjects.
%Among other things a measure is presented, which is known as the right to be forgotten in Art 17 of the GDPR~\cite{SicherheitinderInformationstechnik2013}.
Among other things, especially the right to erasure was introduced in Art. 17 of the GDPR, also known as the right to be forgotten~\cite{SicherheitinderInformationstechnik2013}. 
%The German Bundesverband e.V published its thoughts to the topic Blockchain in connection with the law~\cite{Eichler2018}.
The German Blockchain Association published concerns for the treatment of blockchain technology under the GDPR law~\cite{Eichler2018}.
In addition to suggestions on the interpretation of the current law, they highlight future improvement possibilities.
The right to erasure collides with the concept of the blockchain.
Its immutability is not only not manipulable, but also does not allow intended changes in retrospect.
Such a manipulation would also be desirable to the blockchain to delete illegal content saved on it~\cite{Gibbs2018}. %, but also to fulfill the constraints of the law.

%Scenario logging and auditing
To illustrate the problem we focus on the use case of logging and auditing.
Logging describes the automated recording of system states, product data or other process information.
%In this case, it is independent of the data itself logged in the blockchain.
Thus makes it possible to trace the actions by a computer system for humans. % [ 18 ]. 
Especially in IT security, logging is important to detect weak points and to document security breaches or to investigate them afterwards. %[ 19 ].
However, since log files can also be manipulated, it is mandatory that the authenticity of the log files is given.
Otherwise it can not be used for a reasonable evaluation.
Depending on the scenario, log entries should be stored for different lengths of time.

Based on this, we obtain the following requirements:
\begin{itemize}
\item \textbf{Authenticity}:\\ The act of confirming the truth of an data entry.
\item \textbf{Redundancy of data}:\\ Storing the information multiple times to be fail-safe.
\item \textbf{Delete on request}:\\ A selective erasure of entries or blocks with less effort.
\item \textbf{Scalability}:\\ The system runs for a long time independent of the amount of users and the blockchain size.
 %: very large amounts of data have to be stored depending on the size of the network
\end{itemize}

In the following, we decide between deletion and forgetting of data.
The difference is that deleting is on purpose while forgetting is a delayed procedure.
Nevertheless, if a user of data systems do not follow the rules of erasure, any information published could be stored forever.

\section{Related Work}  %done
%Genesis Block - trust anker

To tackle the problem, different approaches have already been made and other technologies may useable to fulfil the requirements partially.
The most promising techniques are the off-chain transactions,
since interactions with the main chain are comparatively expensive, time-consuming and the whole network has to reach a consensus. %, there are various off-chain solutions.
During off-chain transaction, digital tokens are exchanged and moved without being recorded on the main chain~\cite{bitcoinwiki2017}.
In order to provide guarantee, \textit{Binding Corporate Rules} can be used.
But this approach does not solve the problem for global and public blockchains.
%This just shifts the problem from t
%Even the usage of Off-Chain ledgers technology won't solve the problem.

Another possibility is that not the private user data are stored in the blockchain, but only the hashes of the user data for possible verification.
These could be shared off-chain, while the correctness can still be checked by the hashes.
For example, payment channels and micropayment channels establish a connection between two parties or a closed community.
This can use to perform any number of transactions without being held on the main chain~\cite{bitcoinwiki2018,Poon2017}.
Theoretically, the data of a channel could be forgotten by all participants of the channel deleting the closed channel afterwards.
The main chain is only involved in the beginning and extended with transactions at the end~\cite{Debois2018}.
This also works by storing only asymmetrical encrypted data in the blockchain, while maintaining the keys off-chain.
If the data needs to be retrieved from the ledger, the key has to be requested and the data must be decrypted first.
%The data become effectively erased when the decryption key(s) are not available.
The data can be effectively regarded as erased if the decryption key(s) are not available.
This throttles the transactions per second problem as well as the memory scaling problem.
Nevertheless, this does not fulfil the requirement of reducing the blockchain length, to shrink the data size and erase illegal content.

The approach of Proof-of-Burn can be used to transfer coins or data from one blockchain to another~\cite{P4Titan2014}.
With a complete transfer of all important data, the old blockchain can be deleted.
In context of a GDPR-compliant, it supports the right to erasure with data tokenisation.
In a system, where tokens have to be generated and redeemed for each piece of data in order to process this data, a user requests to burn tokens associated to her account.
Thus, the information embedded in the tokens are no longer usable.
A selective deletion is not possible, especially of not account associated data. Furthermore the effort is very high.

%UltraNote provides self-destructing data storage based on an expiry time) [ UltraNote. Absolute privacy at your fingertips. ].
%However, it is unclear if the underlying cryptomechanisms are usable when erasure is explicitly requested at an arbitrary time

The IBM identity mixer takes a different approach and does not try to disguise or delete the data~\cite{IBM2018}.
It rather disguises the user before even storing the data \cite{Matzutt2018a}.
However, one problem with this approach is that it is no longer possible to assign the data to a person.
This excludes the legal way in case of misconduct.

Furthermore, there are many more approaches, which try to avoid the problem instead of providing an integrated solution.
The simple solution of pruning locally stored parts does not solve the problem for the global, distributed blockchain~\cite{Florian2019}.
Another possibility is a hard fork to a new blockchain~\cite{Ateniese2017} after unwanted content is stored.
But this is very time inefficient as it can take place on every transaction.
Approaches based on chameleon-hash functions leave the responsibility with the key owners and produce a lot effort~\cite{Camenisch2017,Puddu2017}.

All in all, the presented techniques allow to store the data for example in hashed ways or in a way where the data subject is obfuscated.
%Even the usage of Off-Chain ledgers technology will not solve the problem.
No presented solution solves the problem that the data on the distributed blockchain cannot be deleted, and thus the requirements for the use case are not fulfilled.

\section{concept}%done
In this paper, we present the first concept to create the possibility to delete information from a living blockchain system.
This allows the blockchain technology to become a full transactional datastore with high trust and all the other established properties.
Furthermore, this approach ensures that all blockchain systems will remain scalable and manageable even if they run for a long time.

In order to fulfil the requirements, we extend the functionality of the blockchain by creating summary blocks at regular intervals.
Afterwards, we present the concept, which gives the users of the blockchain the possibility to delete their data on request.
This is done after a certain time or with delay.
Therefore, we then describe the process to limit the length of the blockchain and to delete on request.
%In the last step, the weak points of the concept will be addressed.
This is followed by an evaluation, in which we discuss the improvements achieved as well as strengths and weaknesses. 

\subsection{Background and Consensus Algorithm}%done
A direct deletion of an individual entry or a block destroys the hash chain of a blockchain.
This leads to inconsistency, corruption and the loss of trustworthiness.
The information could no longer be completely traced back to the first block anymore.
The first block is called \textit{Genesis Block}.
To be able to delete individual entries on request, the blockchain have to be able to forget or lose information with intent.
Therefore a novel concept is necessary, whereas the action of deletion is approved by the trusted anchor nodes.
These can be seen as the guardians of the blockchain.
For the election of the group of these trusted nodes, several community based approaches can be applied~\cite{Black1958}.
This depends on the type of the blockchain: public, private, consortium, hybrid.
For example, the trusted community could consist of a non-profit organisation or participated users, who have previously done transaction in the blockchain.

One of the main processes of the distributed system is to agree on a common consensus.
The presented concept is based on this functionality, independent of the specific consensus algorithm.
We rely on the outcome of the trustworthy common agreement of the system's nodes for the same knowledge of each nodes.
The agreement on the same blockchain is usually done by some core nodes, called anchor nodes.
These node manage the full copy of the blockchain and build the quorum.
All reliability in a blockchain starts with the common agreement of the quorum on the \textit{Genesis Block} of the blockchain at the beginning.
Among other things, the block number $\alpha$ adresses a specific block in the chain.
It is used to execute a certain function on fixed blocks by each node.
Through the consensus algorithm, it can be assumed that each node has the same level of information~\cite{Hertig2017}.

\subsection{Extension with summary blocks}%done
\begin{figure}[htb]
	%\vspace*{-0.3cm}
	\centering
	\includegraphics[width=0.49\textwidth]{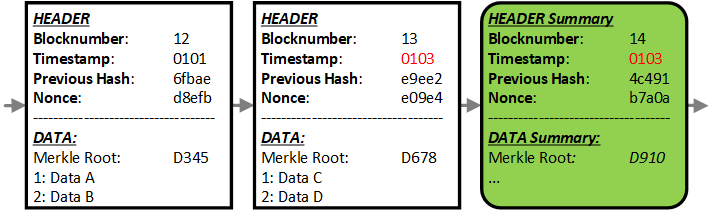}
	\caption{Example of extending the blockchain with a summary block, visualized as green block with round corners.}
	\label{fig:summaryblock}
	%\vspace*{-0.1cm}
\end{figure}

In the first step, the blockchain is extended by a special block type.
Such a block is called summary block $\Sigma$.
It consists of deterministic information only and does not include external or additional information.
A summary block is periodically created by each node itself, for example every tenth block.
These blocks are integrated in the blockchain like normal blocks.
Figure \ref{fig:summaryblock} shows an example of a summary block $\Sigma$.
It can be noted that the block number $\alpha_{\Sigma}$ of the summary block is increased by one as normal blocks.
%This supports the synchronization.
%We just keep it the same to express that it is the summary block for all blocks until this number.
The summary block has the same timestamp $\tau$ as the block before. %, $\alpha_{\Sigma}$ - 1.
This is mandatory so that each node can calculate the summary block with its hash values by itself.

As each anchor node has the same agreed blockchain, the summary block will be identical for each node.
So this information can be used to check synchronisation by comparing the hash of its summary block.
Beneficially, the block do not need to be propagated by itself.
The selected consensus algorithm of a blockchain system has to be extended by this behavior.

%Diskussion
One difficulty with summary blocks is that the nodes must create them themselves and still have exactly the same header.
As all nodes work together with the same consensus algorithms, this should be always the case.
In case of a failure, the hash of the blocks are different, which would result in a fork in the blockchain and thus split the network.

\subsection{Limiting the Sequences in a Blockchain}%done
By inserting the summary blocks, the blockchain can be seen as separated into sequences. % as can be seen in Figure 5.2.
%%This eequences are defined by the summary blocks.
%A sequence $\omega$ is a series of blocks including the summary block at the end of each sequence.
%Regardless of the chosen parameter, the distance between two summary blocks is described by $\delta_{l}$, the length $l$ of a sequence $\omega$.
%A presentation is given in Figure \ref{fig:separatedsequences}, whereas the length $l$ of a sequence $\omega$ can be configured.

\begin{figure}[htb]
	%\vspace*{-0.3cm}
	\centering
	\includegraphics[width=0.49\textwidth]{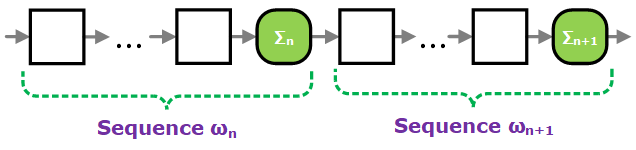}
	\caption{Consideration of the sequences of a blockchain, defined by the summary blocks.}
	\label{fig:separatedsequences}
	%\vspace*{-0.1cm}
\end{figure}

A sequence $\omega$ is a series of blocks including the summary block at the end of each sequence.
The length $l_n$ is defined as the number of blocks in a sequence $\omega_n$.
A representation of the sequences is given in Figure \ref{fig:separatedsequences}.
The lengths $l$ of the individual sequences $\omega_n$ is defined constant depending on the application, but do not have to be.
Regardless of the configurable parameter, the distance between two summary blocks is described by $\delta_{l}$, the length $l$ of a sequence $\omega$..

\begin{figure}[htb]
	\vspace*{-0.1cm}
	\centering
	\includegraphics[width=0.49\textwidth]{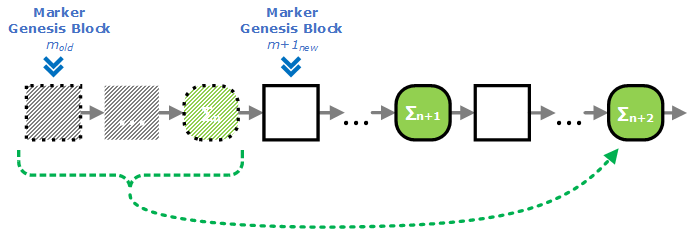}
	\caption{Summarization after exceeding a specified chain length: The first sequence will be deleted after their content is copied into the new summary block.}
	\label{fig:sequenceing}
	%\vspace*{-0.1cm}
\end{figure}

Now, we are able to reduce or shrink the length $l_{\beta}$ of the blockchain $\beta$ and therefore the size of it.
The quorum agrees on a defined blockchain length $l_{max}$. %, more specific on the maximum number of sequences $\omega$, in the living blockchain.
Alternatively, another property can be used, for example the maximum number of sequences $\omega$ in the living blockchain.
If the blockchain grows larger than the specified length $l_{max}$, the oldest sequence will be merged into the next summary block
%So the process starts deleting the sequences from the beginning of the blockchain.
%A sequence older than the defined blockchain size $\beta_{l_{max}}$ 
Therefore, the same information of all single blocks of the first sequence is copied into the new summary block at the end of the chain.
This includes the information of the old summary block to be deleted.
The copied information consists of the data part of the block as well as the block number and timestamp.
The \textit{nonce} and \textit{previous hash} of a block are not needed anymore.
After that, the quorum build a consensus about redefining the \textit{Genesis Block}, based on the blocks of the current chain.
By a majority vote, the quorum determines the new first Block and the time of the changeover.
%The quorum shall take a majority decision.
The new \textit{Genesis Block} will be the block after the summary block of the sequence to be deleted.
A \textit{Marker} $m$ is used to indicate the shifting \textit{Genesis Block}, holding the block number.
As this block is part of the blockchain, it is a trusted anchor for the left blockchain part already approved by the anchor nodes.
Afterwards, the old sequence can be cut off and deleted from the blockchain.
As currently the same information is still in the blockchain, the chain is still valid and complete.
Consequently, this procedure is based on sequences, not on single blocks.
Figure \ref{fig:sequenceing} visualizes the process of summarizing the oldest sequence into the new summary block similar to the round-robin principle.
The green colored and rounded blocks $\sum$ are the summary blocks and the shaded blocks are noted to be deleted.

\begin{figure}[htb]
	%\vspace*{-0.3cm}
	\centering
	\includegraphics[width=0.42\textwidth]{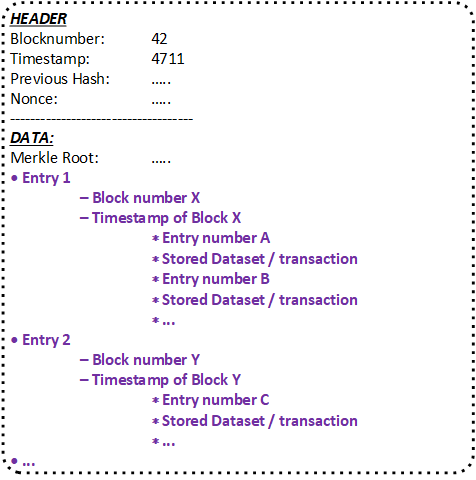}
	\caption{Data structure of the summary blocks.}
	\label{fig:blockstructure}
	%\vspace*{-0.1cm}
\end{figure}

The data part of a summary block has the following structure as in Figure \ref{fig:blockstructure}, whereas the content is copied from the original blocks.
So, the block number, the timestamp and the entry number are keeped the same as initially integrated.

%\begin{itemize}
%\footnotesize
%\item Entry 1
%\begin{itemize}
%\item Block number \textit{X}
%\item Timestamp of Block \textit{X}
%\begin{itemize}
%\item Entry number A
%\item Stored Dataset / transaction
%\item Entry number B
%\item Stored Dataset / transaction
%\item ...
%\end{itemize}
%\end{itemize}
%%\begin{itemize}
%\item Entry 2
%\begin{itemize}
%\item Block number \textit{Y}
%\item Timestamp of Block \textit{Y}
%\begin{itemize}
%\item Entry number C
%\item Stored Dataset / transaction
%\item ...
%%\end{itemize}
%\end{itemize}
%\end{itemize}
%\end{itemize}

\subsection{Deleting information on request}%done
To enable the deletion of specific information, a participant needs to be granted the additional functionality to send a delete request.
%Therefore, a client submits a delete request in form of a delete entry or delete transaction beside the normal entries to the blockchain system.
For this purpose, a user submits such a request to the blockchain system in form of a deletion entry or a deletion transaction.% in addition to the normal entries.
This follows the same procedure as normal entries and transactions.
%To do this, clients must create a transaction that requests deletion.
The associated data set to be deleted is referenced by the block number and the according entry number, in which the data set is stored. % followed by the consecutive numbering given to the data entry.
It may happen that an entry is located in a summary block. This must be taken into account by also considering the data records in summary blocks.
The complexity of the procedure is linear and very low as a blocks are references directly by number.
Regarding the deletion of information, the following two aspects of authorization and semantic cohesion need to be considered.
Furthermore, the challenge of delayed deletion and the possibility of temporary entries is addressed.

\subsubsection{Authorization}%done
To ensure that the user is authorized to have the information deleted, a deletion request must be signed with the client signature just like a normal entries.
%To make sure that no participant can delete any information, a delete request has to be signed with the client signature as normal data sets.
For authorization of privileges, it can be applied a role-based concept with corresponding user signatures.
%Applying a role-based concept or the users, authorization is based on the corresponding role signatures.
%For the authentication of privileges, a user-based role concept is applied.
%The authentication takes place using appropriate role signatures.
Particular reference is made to authentication approaches based on blockchains~\cite{Nagpal2018, Mohsin2019}.
For example, the anchor nodes of the quorum work together as a basis of trust and are jointly granted full administrative privileges.
%For example, the anchor nodes of the quorum as trust base work together and obtain full administrative privileges in common.
These receive a master signature.
Based on a protocol for consensus-based voting, the quorum dictates the requirements and constraints for edition of the editing~\cite{Deuber2019}.
In comparison, a user is only allowed to submit delete requests for his own transactions.
These can be identified by comparing the signature of the user and the stored signature of an data entry.
The system checks if the signatures share the same key and the privileges are given according the role-based concept.
%Before a deletion request is approved, 
According to the consensus of the anchor nodes, a deletion request is approved and entered in the blockchain.

\subsubsection{Semantic Cohesion}%done
The semantic correctness of the complete blockchain has to be proven by the quorum before a deletion takes place.
A deletion request can only be granted, if further transactions do not rely on it.
Otherwise, multiple transactions need to be revoked, which may involves additional parties.
A deletion request of such a chain part of a transaction chain can be approved by the signatures of all dependent parties.
So these entries become orphaned in the meaning of the content \cite{Nandi2018}.
At this point, the quorum must ensure that the cohesion in the blockchain is maintained and remains valid after the deletion.
This depends strongly on the use case and the specific scenario.
Therefore, a deletion request has to be proven manually.
An automatic approached could be designed based on the principle of \textit{Bell-LaPadula model} or \textit{Brewer-Nash Model}.

\subsubsection{Delayed Deletion}%done
In this case, deletion means that the information is not deleted immediately. 
For the moment, the specified data is marked to be deleted in the future.
Subsequent incoming transactions based on this marked data are no longer permitted.
%Deletion in this case means that the information is not deleted immediately.
%Currently, the specified data are marked to be deleted in the future.
%Additional incoming transaction based on these marked data to be deleted are not allowed anymore.
A marked data set will be deleted when the sequence gets too long or too old, see Equation~\ref{eq:shrink}.
This is the case when the marked data reaches the beginning of the blockchain.
As a result, the sequence is considered old enough and the information to be deleted is no longer included in the corresponding summary block.
%This means it reaches the beginning of the blockchain.
%After that, the sequence is old enough and the information is no longer included in the corresponding summary block.
%
\begin{equation}
\label{eq:shrink}
\begin{split}
if\; l_{\beta}\; >\; l_{max},\; then\; l_{\beta_{new}}\; =\; l_{\beta_{old}}\; -\; l_{\omega_{1}},\\(whereas\; \omega_1\; is\; the\; beginning\; of\; \beta)
\end{split}
\end{equation}

\begin{figure*}[htb]
	%\vspace*{-0.3cm}
	\centering
	\includegraphics[width=0.98\textwidth]{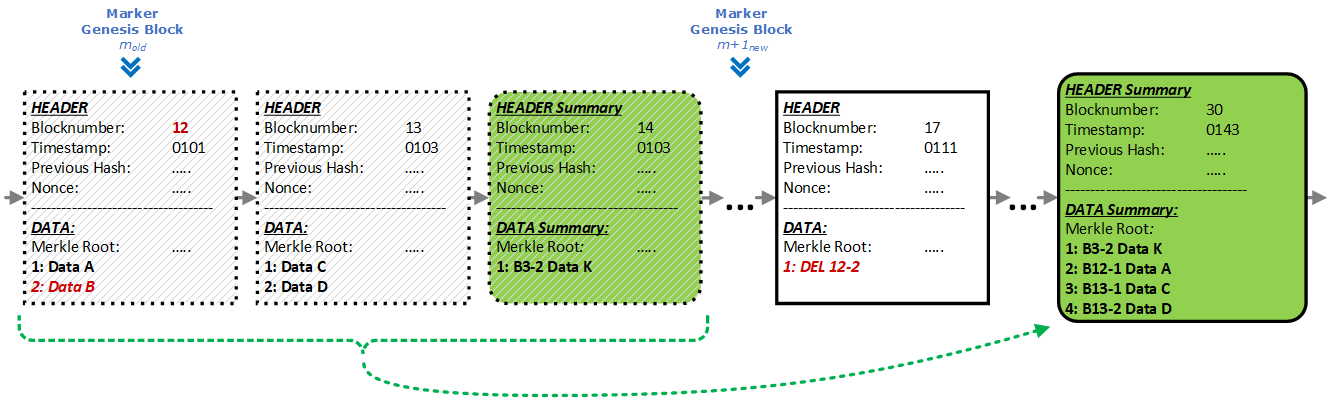}
	\caption{Selective deletion on request. The colored marked entry is not copied into the new summary block due to the delete request. The first
		sequence will be deleted after their content is copied.}
	\label{fig:summaryblockdeleted}
	%\vspace*{-0.1cm}
\end{figure*}

Figure \ref{fig:summaryblockdeleted} shows how an entry is deleted with delay by the blockchain using a deletion request.

Since there is no need to keep deletion requests longer then necessary they will never be copied into a summary block.
To prevent a long delay in deletion, a possibility is to extend the blockchain with empty blocks.
This can be done, by regularly adding empty blocks after a time interval if no transaction has occurred.
%Furthermore, the all delete request are also not copied to the summary block.

To avoid shortening the blockchain too much, a minimum length or a minimum number of summary blocks can be specified for the remaining blockchain.
Another criterion for ensuring that sufficient data remains in the blockchain is a minimum time span coverage.
%Another criteria to ensure to keep the data in the blockchain is a minimum time span coverage.
So, there are two possibilities to determine the length of the blockchain: by time and by (summary) block quantity.
New sequence in a blockchain may be created, although the oldest sequence is not yet old enough to be deleted.
This creates empty summary blocks.
This case for example occur at the beginning of the blockchain.
However, multiple sequences can also being combined in one summary block. 

\subsubsection{Temporary entries}
The current design can be extended by temporary entries, which will be deleted automatically after a specific timestamp ($\tau$) or block number ($\alpha$) has been reached.
Therefore, an entry is extended by an optional data field in which the maximum storage time is recorded (e.g. 42: Data T : $\tau$8888 or $\alpha$4711).
If the blockchain exceeds the timestamp or block number given, the entry will not be transferred to the new summary block.
So it is automatically removed from the blockchain without additional authorization needed.
The system cleans up its own content and deletes the temporary information.
%The system clears up its content on its own and deletes information that is too old after time has been elapsed.
Use cases for this functionality include log files of operating systems or products traced in a blockchain according to Industry 4.0~\cite{Banerjee2018}.
%A typical use case for this functionality are log files of operating systems or products traced in a blockchain according to Industry 4.0~\cite{Banerjee2018}.
%Quelle: https://www.infosys.com/Oracle/insights/Documents/product-tracking-tracing.pdf

\section{Evaluation} %done
To evaluate our concept for selective deletion in a blockchain, we have implemented a prototype of the system.
The implementation is realized as a client-server system using Python and Java.
In order to be language independent and to get reusable components, the middleware technology of CORBA is used.
%The concept is realized prototypically in Python and Java as Client-Server System.
%Therefore the middleware technology of CORBA is used to be implementation language independent and to obtain reusable components.
The blockchain can be seen as a list of blocks maintained by server nodes, namely the anchor nodes.
If the sequence reaches its defined length, a summary block will be created automatically.
This includes the check about blocks to be deleted.
%Entries from the summary block do not need to be parsed for delete  and will be taken over directly to speed up the process.
%Whereas entries to be forgotten will be excluded from the new summary block. 

To evaluate the concept, we focus on the use case of logging as mentioned in Section \ref{sec:scenario}.
All logins to a terminal are logged to the blockchain.
Therefore, the signature of each specific user login is stored in a block.
In this way, the authentication of the user can be monitored and audited.
The blockchain system is configured to create a summary block for every third block.
%The blockchain system is configured in that summary blocks are created every third block.
The persons in the scenario are referred to as \textit{ALPHA}, \textit{BRAVO} and \textit{CHARLIE}.
%A persons in the scenario are named after \textit{ALPHA}, \textit{BRAVO} and \textit{CHARLIE}.
%Whereas, the user \textit{CHARLIE} has Administrator priviledges.
In the following test setup, the specific signatures of each client prevent a user from deleting an entry of another user.
%In the following test, the specific signatures of each client prevents a user from causing an entry from another user to be deleted.
% The blockchain has a length of 60 seconds and stores the information forever unless it requested to be deleted.

%For better understanding the concept is somewhat simplified in the implemented prototype.
To visualize the blockchain, the entries are listed line by line.
Each block has the following header structure: block number; timestamp; previous block hash; own block hash; optional data entry.
An data entry is structured as follows: D stores data record; K holds the user; S poses as signature (here simplified).
The structure of a data entry is specified beforehand by a YAML schema.
For simplification, blocks starting with S are the summary blocks.
These are created automatically and marked in blue.

\begin{figure}[htb]
	%\vspace*{-0.3cm}
	\centering
	\includegraphics[width=0.49\textwidth]{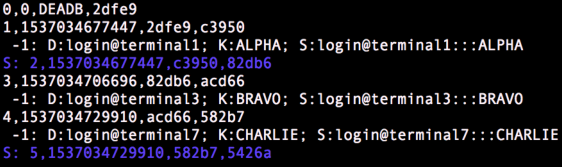}
	\caption{Console output shows the current status of the content in the block after three logins.}
	\label{fig:example1}
	%\vspace*{-0.1cm}
\end{figure}

Figure \ref{fig:example1} shows the \textit{Genesis Block} 0 with previous hash \textit{DEADB}.
The first two summary blocks are empty, because there was nothing to add up from previous sequences yet.
The maximum length of the blockchain is not reached, so nothing gets deleted.
Each user has an entry in block 1,3 and 4.

\begin{figure}[htb]
	%\vspace*{-0.3cm}
	\centering
	\includegraphics[width=0.49\textwidth]{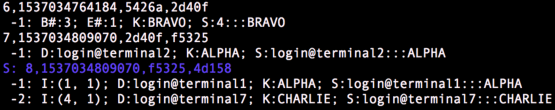}
	\caption{The first two sequences are summarized into the latest summary block, whereby the entry to be deleted was not copied.}
	\label{fig:example3}
	%\vspace*{-0.1cm}
\end{figure}

Figure \ref{fig:example3} shows a request on deletion in block number~6.
The user \textit{BRAVO} requests the deletion of its entry in block number~3 and data entry~1.
Since the issuer of the entry to be deleted matches the user of the deletion request, the corresponding entry is not transferred to the summary block at the next action of deletion.
Furthermore, the first and second sequence of the blockchain is merged into the last summary block.
The entry from block number~3 is not copied.
The maker for the \textit{Genesis Block} is changed to block number~6.
All information before block~6 is deleted.

Nevertheless, wrong request of deletions can be included in the blockchain, but these have no further effects.

\begin{figure}[htb]
	%\vspace*{-0.3cm}
	\centering
	\includegraphics[width=0.49\textwidth]{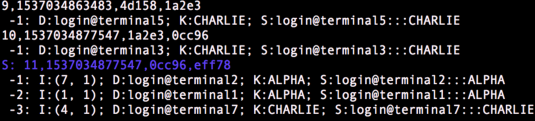}
	\caption{Console output one cycle ahead without deletion request in the summary block.}
	\label{fig:example4}
	%\vspace*{-0.1cm}
\end{figure}

Figure \ref{fig:example4} presents the blockchain one cycle ahead.
The deletion request entry of \textit{BRAVO} is no longer stored in the current blockchain, as deletion entries are never transferred.

%Datensätze zusammenführen, Aufsummierte Transactionen speichern und nicht nur einzelne Daten.
\subsection{Achieved Enhancements} %done
Our concept has multiple advantages, offers additional possibilities, and solves the growing data problem:
\begin{itemize}
%\item The block header information can be omitted.
\item \textbf{Data Reduction}:\\ Deleting and forgetting unwanted content by not copying the information to the new summary block.
\item \textbf{Corrections}:\\ Change information, which maybe submitted wrongly.
\item \textbf{Reorganization}:\\ Possibility to reorganize information for better scalability.
\item \textbf{Summarized Information}:\\% If A makes a transaction to B and B a further transaction to C, only the results that A makes a transaction to C can be stored in the blockchain.
The ability to summarize coherent information. E.g., if a system logs an event several times, these logs can be stored summarized in the blockchain.
\item \textbf{Recovery}:\\ In the case of cryptocurrencies, it offers the possibility to make lost coins usable again. It means not for a single user, but for the entire blockchain system to prevent a system shutdown in long term~\cite{Roberts2017}.
\end{itemize}

\subsection{Discussion: Strengths and Weaknesses} %done
To realize the possibility of deletion in a blockchain, certain trade-offs are made.
%In order to realize the possibility of deletion in a blockchain there are certain trade offs that are made.
\subsubsection{Hamper 51 \% attack}
\begin{figure}[htb]
	%\vspace*{-0.3cm}
	\centering
	\includegraphics[width=0.49\textwidth]{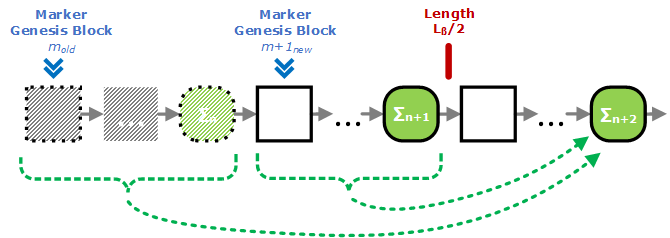}
	\caption{Hampering the 51 \% attack by storing data more than once in the blockchain}
	\label{fig:hampering}
	%\vspace*{-0.1cm}
\end{figure}

%\textit{\newline\hspace{0.3cm}1) Hamper 51 \% attack:\\}
The most known attack on a blockchain is the 51 \% attack \cite{Madore2019}, where an attacker controls more than 50 \% of the resources to create new blocks or manipulate the block order.
The precise amount depends on the consensus algorithm.
This attack is also the base for enhanced attacks such as double spending and feather forking.
During the process of a blockchain, each newly added block is a confirmation of all previous blocks.
Since the old sequence in which the information was stored is deleted, there is no possibility for new nodes to validate the correctness of the most current summary block.
To counter this possibility, we make use of the summary blocks to store data more than once, or at least the Merkle root as reference for validity to reduce the amount of data.
Therefore, the sequence to be deleted and the reference to a middle sequence, for example $\omega_{l_{\beta} / 2}$, is stored in the new summary block, see Figure~\ref{fig:hampering}.
So, each entry that is longer than $l_{\beta} / 2$ in the blockchain has at least $l_{\beta} / 2$ confirmations at each time.
The trustworthiness is given, even if the oldest entry is combined in the latest summary block,
This hampers the 51 \% attack against a blockchain without summary blocks as for an attacker it is not enough to run the attack for only one block.
The attacker has  to run the attack for a least $l_{\beta} / 2$  number of blocks.
The longer the blockchain is and the more summary blocks there are, and consequently the more trustworthy the chain is.

\subsubsection{Size of summary blocks}
By adding up the information in summary blocks, they become larger over time.
The creation of these summary blocks can take a long time, depending on the amount of data to be copied and the storage performance.
One possibility to counter this is to structure the information logically and build packages.
In addition, the copying of much information can be avoided by working with hash references.
The data packets are stored separately and only linked in the blockchain, as with other off-chain approaches.
Another opportunity is to summarize the content as often the result is mandatory to be stored, not every single transaction.

\subsubsection{Extending Consensus algorithm}
Any consensus algorithm can be extended by the described behavior.
Nevertheless, it needs to be ensured that the participants do not evaluate the blockchain based on the length or the value of the block index as the chain can be shortened.
This can be prevented if the nodes only accept a blockchain which is traceable from its current status quo.

\subsubsection{Isolating nodes or evil nodes}
Node isolation is an attack where an attacker tries to separate participants from the network by controlling partner nodes of the target.
Known attacks are Eclipse attack \cite{Heilman2015} or Sybil attack~\cite{Douceur2002}.
Thus, an attacker can weaken or completely destroy the network piece by piece.
Therefore, the blockchain system has to have some anchor nodes, whereas clients obtain the current status quo of the blockchain.

\subsubsection{Ascertaining the quorum and reaching a consensus}
The problem of quorum building is a long known problem in computer science.
There are different methods to solve this problem, especially for the selection of anchor nodes for distributed systems~\cite{Skeen1982}.
Depending on the type of the blockchain, these procedures offer different advantages and disadvantages.
A decision has to be made individually.
This also applies to the selection of a suitable consensus algorithm.
The presented approach in this publication is independent of the concrete characteristics of quorum selection and consensus algorithm.

\section{Summary}%done
In this paper, we presented the first concept to delete entries in a blockchain while maintaining the characteristic attributes like trustworthiness and consistency.
%The presented concept allows the deletion of entries in a blockchain, while maintaining the trustworthiness.
In particular, the requirements of laws to delete personal data can be met such as defined in the European Data Protection Act~(GDPR).
In summary, the concept is based on an extension of the behavior of the consensus algorithm.
The blockchain is divided into sequences by regularly adding summary blocks.
These summarize the old information of the first sequences and store the data at the current end of the blockchain.
During this process, invalid, obsolete, or to be deleted entries are ignored.
Afterwards, the marker for the \textit{Genesis Block} is shifted to the following sequence.
The first sequence can be forgotten and will be deleted.
So, the blockchain only contains data that is relevant for the current and future time.
This solves the problem of a growing blockchain size.
Furthermore, the blockchain technology becomes full transactional.
A deletion request requires a certain amount of time to be processed, which can be seen as a disadvantage.
But it is mandatory to avoid illegal manipulations.
In the end, the system is not as dynamic as classical database systems.
Nevertheless, the block structure as well as the network are independent of this solution.
Also, the added functions are independent of the consensus itself and can be combined with different strategies.
So, the approach is flexible adaptable to extend all blockchain systems as well as already running systems like Bitcoin.

Typical fields of application are not limited to the area of cryptocurrencies to cope with the increasing size of the blockchain.
In the area of Industry 4.0, the production of a good can be recorded along the entire supply chain.
As soon as the minimum best-before date has been exceeded or the data has expired, the new technology can be used to automatically clean up the blockchain.
In the field of vehicle maintenance, the life cycle of each car can be documented centrally, so that manipulations are excluded, e.g. on the milage or accidents.
After a vehicle is taken out of service, the blockchain as database is cleaned up to handle the data amount.
In the future, we will implement the concept in upcoming blockchain frameworks like Hyperledger, Ethereum or the Enterprise blockchain: MultiChain.

\section*{Acknowledgment}%done
Special thanks goes to Simon Ostendorff for the development of the basic prototype and his passionate support for this research.

\bibliographystyle{unsrtnat}
\bibliography{IEEEexample} 

\end{document}